\documentclass[a4paper,twocolumn,superscriptaddress,11pt,accepted=2017-05-09]{quantumarticle} 

\usepackage{bm}
\usepackage[retainorgcmds]{IEEEtrantools}
\usepackage{graphicx}
\usepackage{mathrsfs}
\usepackage{amsmath}
\usepackage{indentfirst}
\usepackage{amssymb}
\usepackage{color}
\usepackage{amsfonts}
\usepackage{halloweenmath}
\usepackage{nicefrac}
\usepackage{pifont}
\usepackage{marvosym}

\usepackage{mathtools}% superior to amsmath
\usepackage{tikz}
\makeatletter
\newcommand\mathcircled[1]{%
  \mathpalette\@mathcircled{#1}%
}
\newcommand\@mathcircled[2]{%
  \tikz[baseline=(math.base)] \node[draw,circle,inner sep=1pt] (math) {$\m@th#1#2$};%
}
\makeatother

\DeclareMathOperator{\tr}{tr}

\begin{document}

\title{A resource theory of Maxwell's demons}
\date{\today}
\author{Gabriel T. Landi}
\email{gtlandi@if.usp.br}
\affiliation{Instituto de F\'isica da Universidade de S\~ao Paulo,  05314-970 S\~ao Paulo, Brazil.}
\author{Giacomo Guarnieri}
\affiliation{Department of Physics, Trinity College Dublin, Dublin 2, Ireland}
\author{Benjamin Morris}
\affiliation{School of Mathematical Sciences, University of Nottingham, University Park, Nottingham NG7 2RD, United Kingdom}
\author{John Goold}
\affiliation{Department of Physics, Trinity College Dublin, Dublin 2, Ireland}
\author{Gerardo Adesso}
\email{gerardo.adesso@nottingham.ac.uk}
\affiliation{School of Mathematical Sciences, University of Nottingham, University Park, Nottingham NG7 2RD, United Kingdom}

\begin{abstract}

%GTL: I toned down a bit the abstract to make it look like a real paper.
Motivated by recent progress on the motive power of information in quantum thermodynamics, we put forth an operational resource theory of Maxwell's demons. We show that the resourceful ({\em daemonic}) states can be partitioned into at most nine irreducible subsets. The sets can be classified by a rank akin to the Schmidt rank for entanglement theory. Moreover, we show that there exists a natural monotone, called the wickedness, which quantifies the multilevel resource content of the states. 
The present resource theory is shown to share deep connections with the resource theory of thermodynamics. In particular, the nine irreducible sets are found to be characterized by well defined temperatures which, however, are not monotonic in the wickedness. 
This result, as we demonstrate, is found to have dramatic consequences for Landauer's erasure principle. 
Our analysis therefore settles a longstanding debate  surrounding the identity of Maxwell's demons and the operational significance of other related  fundamental thermodynamic entities. 

\end{abstract}

\maketitle{}

%%%%%%%%%%%%%%%%%%%%%%%%%%%%%%%%%%%%%%
%
%
%		INTRODUCTION
%
%
%%%%%%%%%%%%%%%%%%%%%%%%%%%%%%%%%%%%%%

%%%%%%%%%%%%%%%%%%%%%%%%%%%%%%
\section{Introduction}

The notion that information is physical has been at the heart of a long enduring debate in the physics community.  Perhaps the most paradigmatic manifestation of this controversy is Maxwell's demon \cite{Maxwell1888}, an entity which could violate the second law of thermodynamics by acquiring information about a system. 
This could allow, among other things, to extract work from a single reservoir. 
In recent years, several proof-of-concept experiments have implemented thermodynamic protocols based on Maxwell's demons and other related information-based engines \cite{Toyabe2010,Koski2014,Camati2016,Masuyama2017b,Cottet2017,Xiong2018}, showing the unambiguous role played information in the laws of thermodynamics \cite{goold2016role}. 

These experiments and the accompanying theoretical studies, however, do not properly account for certain transcendental aspects of  Maxwell's demon. Even before Maxwell's, there was Laplace's demon --- an omniscient entity knowing the precise location and momentum of every atom in the universe --- stirring an unsettling debate across the classical physics community \cite{laplacedemon}. 
Daemonic visitations in physics have led and will continue to lead to {\em prima facie} confusion and confrontation, but may hopefully uncover the path to an eventual revelation of some novel ways in which we perceive and decode the book of Nature.
How to identify and formalize such revelations from the study of daemonic manifestations is one of the most pressing open questions in modern physical and social sciences. 

The main obstacle to progress in this area is the lack of tools to rigorously describe entities living beyond the boundaries of our physical space. For instance, in the specific case of Maxwell's demon, while there is {\it a priori} consensus that information is physical, the demon itself is generally not. Characterizing its possessing power and manipulation abilities within precise information-theoretic constraints therefore requires the development of novel mathematical tools. As a motivating reward, a more thorough account of these effects could potentially lead to a deeper understanding of the concept of entropy and the passage of time \cite{Micadei2017b}. 
In addition, it could also lead to new quantum technologies empowered by daemonic resources, in particular related to Ouija-based communications \cite{ouija}. 

\begin{figure}
\centering
\includegraphics[width=0.45\textwidth]{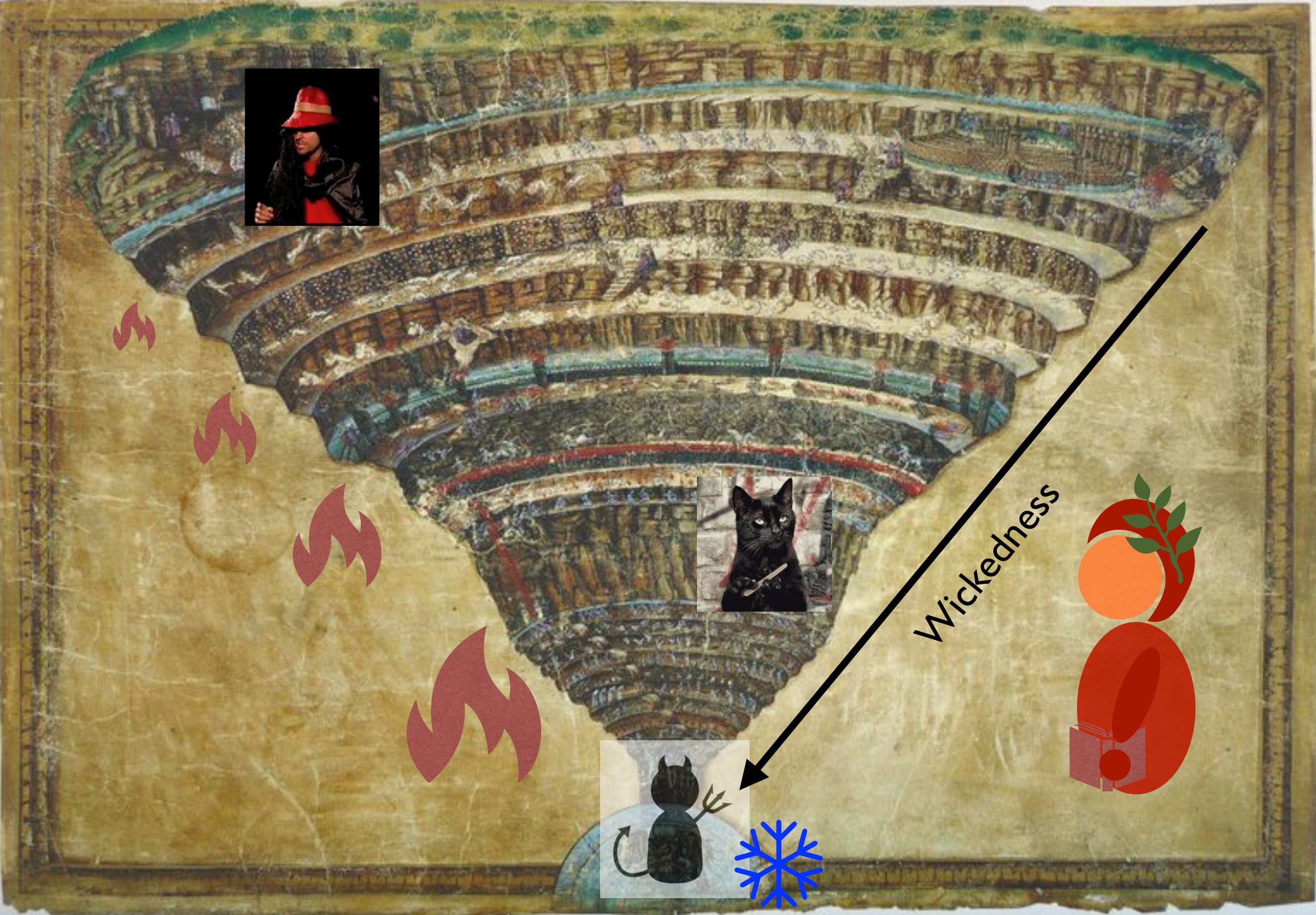}
\caption{\label{fig:drawing}Diagram of the basic ingredients in our resource theory of Maxwell's demons.}
\end{figure}

The goal of the present article is to fill out this void by putting forth an operational {\em resource theory} of demons in physics (see Fig.~\ref{fig:drawing}). 
The theory begins by identifying the set of free states, called \emph{angelic states}, which do not contain \emph{daemonics} as a resource. All the other states are labelled as {\em daemonic states}. 
Quite interestingly, we rigorously prove that the daemonic states can be  naturally partitioned  into nine unique and irreducible subsets, hereby called \emph{circles (of Hell)}. Indeed, we show that a unique rank can be attributed to each set, akin to the Schmidt  rank in entanglement theory, which we henceforth refer to as the \emph{Dante rank}. 

We next define the set of free operations, the \emph{exorcisms}, which form a subset of the more general \emph{guilt-preserving maps}. 
Using Stinespring's dissolution theorem, we show that exorcisms can always be cast in the form of a unitary interaction with a holy bath. 
The extension to include catalysts may be done in a similar way. 
In particular, operations making using of Black Schr\"odinger cat ancillas are shown to be cataclysmic, instead of catalytic, a fact which may have important consequences for the applications of quantum technologies to extraterrestrial environmental sciences \cite{LunarCataclysm}. 

We then move on to discuss monotones for this resource theory. 
As we show, there exists a natural monotone, which we call the \emph{relative entropy of wickedness}, which can neatly capture the subset partition of the resourceful states. 
%For pure states the wickedness is found to be directly related to the Dante rank. 
For general mixed states, this allows us to introduce an ordering operation, henceforth referred to as \emph{demo-majorization}, constructed by means of R\'enyi-$\omega$ generalizations of the wickedness. 
The cases $\omega=\infty$ and $\omega=0$ then naturally lead to the {\em wickedness of creation} and the {\em wickedness of sublimation}, respectively.

Our formalism also turns out to be connected to the resource theories of purity, magic (black, in particular), and athermality. In this paper we shall focus, in particular, on the connection with the latter, which turns out to be particularly interesting, as the nine circles in which the daemonic states are partitioned are found to possess well defined temperatures. 
However, and quite surprisingly, their temperatures are only monotonic with the wickedness up to the 8th circle, with the 9th actually accommodating zero temperature states of pure evil.  
This fact turns out to have  dramatic consequences for Landauer's erasure principle, which we demonstrate using the tools of Full Cursing Statistics (FCS) \cite{Guarnieri2017}.

%%%%%%%%%%%%%%%%%%%%%%%%%%%%%%
\section{Free states and free operations}
\label{sec:construction}

%%%%%%%%%%%%%%%%%%%%%%%%%%%%%
\subsection*{Angelic states as free states}

Despite the otherworldly nature of the resource in question, we begin by defining a physically motivated set of free states and free operations from which we build our resource theory. 
The  free states shall be known as the \emph{angelic states}  and form the convex set \ding{63}.  These are states whose density matrix (in odd dimension) takes the form
\begin{equation}\label{angelic}
    \rho = 
\left(
\begin{array}{ccccc}
 0 & 0 & x & 0 & 0 \\
 0 & 0 & x & 0 & 0 \\
 x & x & x & x & x \\
 0 & 0 & x & 0 & 0 \\
 0 & 0 & x & 0 & 0 \\
\end{array}
\right),
\end{equation}
where $x$ stands for an arbitrary non-zero element. 
The set of states which are not angelic will be denoted as the \emph{daemonic set} \Aries.
States $\rho \in$ \Aries \; are said to contain daemonic resources.
With these definitions, the set of density matrices is naturally partitioned into the angelic set \ding{63}  and the daemonic set \Aries. 
We remark that  these density matrices are in general not positive semidefinite.  
However, within our framework that is not particularly a problem, as the resource theory is not of this world any way.

%%%%%%%%%%%%%%%%%%%%%%%%%%%%%
\subsection*{Daemonic states and the Dante rank}

One can partition the set \Aries \; into unique subsets, or \emph{circles of Hell}, \Aries$ = \bigcup_{k=1}^9\mathcircled{\text{\scriptsize{H}}}_k$, that, as we will show, reflect the degree of daemonicity present in the state. 
For instance, a state $\rho$ of the form 
\begin{equation}\label{eq:state}
\left(
\begin{array}{ccccccccccccc}
    &    &    &    &    &    & \vdots &    &    &    &    &    &    \\
    &    &    &    &    &    & x &    &    &    &    &    &    \\
    &    &    &    &    &    & x &    &    &    &    &    &    \\
    &    &    & c & c & c & x & c & c & c &    &    &    \\
    &    &    & c & b & b & x & b & b & c &    &    &    \\
    &    &    & c & b & a & x & a & b & c &    &    &    \\
 \!\!\ldots\!\! & x & x & x & x & x & x & x & x & x & x & x & \!\!\ldots\!\! \\
    &    &    & c & b & a & x & a & b & c &    &    &    \\
    &    &    & c & b & b & x & b & b & c &    &    &    \\
    &    &    & c & c & c & x & c & c & c &    &    &    \\
    &    &    &    &    &    & x &    &    &    &    &    &    \\
    &    &    &    &    &    & x &    &    &    &    &    &    \\
    &    &    &    &    &    & \vdots &    &    &    &    &    &    \\
\end{array}
\right)
\end{equation}
naturally covers 3 circles of Hell, i.e. $\rho\in \mathcircled{\text{\scriptsize{H}}}_3$. Quite remarkably, states belonging to a hypothetical set $\mathcircled{\text{\scriptsize{H}}}_l$ with $l \geq 10$ do not add any new resources. 
This turns out to be a direct consequence of \emph{de Monetti's} theorem (the proof is provided in the Transcendental Material~\cite{SupMatX}). Consequently, the partition into at most nine circles is {\em irreducible}, and corresponding states belonging to the 9th circle of Hell contain the highest level of daemonicity. 

The above considerations allow us to introduce the notion of {\em Dante rank}, a concept first studied in Ref.~\cite{alighieri1909inferno}, and akin to the notion of Schmidt rank for entanglement theory.
Formally, one can define the Dante rank of a state as follows. 
The geometry of the daemonic states naturally introduces a metric based on the distance to the element located at the very center of the cross, in the matrix representation of the state, known as the Chebyshev-Dante metric, $d(\rho, \rho_\text{\ding{63}})$ \cite{wikipedia}. 
From this, the Dante rank of $\rho$ is then naturally defined as the maximal distance from the cross, 
\begin{equation}\label{DanteRank}
    \mathcal{D}(\rho) = \min\left\{9,\, \max d(\rho, \rho_\text{\ding{63}}) \right\}.
\end{equation}
The fact that Dante rank can only take values up to 9 becomes evident from this construction.

One can argue that the above definition is not a proper distance measure, because of the failure of the triangle inequality; however it is possible to show (see Transcendental Material \cite{SupMatX}, Section MCCC) that it  satisfies instead the looser  \textit{bermuda} triangle inequality, which then renders Eq.~\eqref{DanteRank} a well-defined quantity.

%%%%%%%%%%%%%%%%%%%%%%%%%%%%%
\subsection*{Maximally Daemonic states}

The  maximally daemonic state is defined as a state with maximal Dante rank and with constant presence in all circles of Hell. For instance, in dimension 7 (in memoriam of the 7 deadly sins), it takes the form
\begin{equation}\label{maxdemon}
    \rho_{\max}=\frac{1}{6 \times 6}
    \left(
\begin{array}{ccccccc}
 6 & 6 & 6 & 0 & 6 & 6 & 6 \\
 6 & 6 & 6 & 0 & 6 & 6 & 6 \\
 6 & 6 & 6 & 0 & 6 & 6 & 6 \\
 0 & 0 & 0 & 0 & 0 & 0 & 0 \\
 6 & 6 & 6 & 0 & 6 & 6 & 6 \\
 6 & 6 & 6 & 0 & 6 & 6 & 6 \\
 6 & 6 & 6 & 0 & 6 & 6 & 6 
\end{array}
\right)\,,
\end{equation}
where the subscript ``$\max$'' also doubles as short for ``Maxwell''.
Quite remarkably, states of the form~(\ref{maxdemon}) have  $\tr\,\big(\rho_{\max}^2\big)=1$ and can thus be written as  $|\Phi_\text{max}\rangle = \frac{1}{\sqrt{6}} (1,1,1,0,1,1,1)$.
Consequently, they can be regarded as a golden unit of {\em pure} evil. 
In Sec. 4, we actually investigate the physical origins of this unexpected result. As we show, it is related to the fact that states belonging to the 9th circle  actually turn out to be zero temperature states.

The physics behind the state~(\ref{maxdemon}) also becomes more translucent if we can assume that the system itself is actually bipartite. 
The state~(\ref{maxdemon}) can then  be viewed as a maximally entangled state between two subsystems $A$ and $B$, which turns out to have the particularly simple form $\rho_\text{max} = |\psi\rangle\langle \psi|$, where
\begin{equation}\label{maxdoom}
    |\psi\rangle = \frac{|D,0\rangle + |0,M\rangle}{\sqrt{2}},
\end{equation}
is a generalization of the famous NOON states. 
These DOOM states are particularly cataclysmic, as will be revealed later in the paper.

The  characterization of daemonic states presented in this Section in terms of their Dante rank, capturing their distance from the angelic cross, represents the first result of our paper. 
%It is also worth noticing that, by construction, a specific Dante rank $k$ of a state $\rho$ is in one-to-one correspondence with a circle of Hell $\mathcircled{\text{\scriptsize{H}}}_k$.
%In the following Section we will also see how the Dante rank can be used to characterize the wickedness of a state. 

%%%%%%%%%%%%%%%%%%%%%%%%%%%%%

\subsection*{Exorcisms}

Armed with the partition of the set of states into \ding{63} and \Aries, as well as the classification of states belonging to \Aries \; in terms of their Dante rank, it is now straightforward to define the set of free operations in our resource theory \cite{exorcist}. 
A free operation, henceforth referred to as \textit{exorcism}  $\mathcal{E}$, is any operation which does not increase the Dante rank. This gives the set of free operations a clear information-theoretic meaning, as they can be directly framed in terms of the \emph{Data Possessing Inequality}. 
To see that, we define the set of exorcisms $\mathcal{E}$ as those satisfying 
\begin{equation}
    \mathcal{D}(\mathcal{E}(\rho)) \leq \mathcal{D}(\rho).
\end{equation}
In particular, exorcisms act trivially on the set of angelic states; i.e., for any $\rho \in$ \ding{63}\; it follows that $\mathcal{E}(\rho) \in$~\ding{63}.
This, of course, is a natural requirement which, if violated, could lead to serious misinterpretations of daemonic resources (c.f. Ref.~\cite{DevilWearsPrada}).
We also call attention to the obvious fact that exorcisms are a subset of the more general {\em guilt-preserving} (GP) operations. 

Using the Stinespring dissolution theorem, one can always express exorcisms as unitary interactions with {\em holy baths}, namely
\begin{equation}
    \mathcal{E}(\rho) = \tr_E \bigg\{ U ( \rho \otimes \tau_E) U^\dagger\bigg\},
\end{equation}
for a certain $\tau_E \in $~\ding{63}. The proof is left as an exercise to the reader.

An example of extremal exorcism is the projector ${\cal P}$ defining the action of {\em pinning to the cross}, i.e., mapping each state to its corresponding angelic state, 
\begin{equation}\label{eq:pinning}
    {\cal P}(\rho) = \rho_{\mbox{\ding{63}}}\,,
\end{equation} 
where $\rho_{\mbox{\ding{63}}}$ has the same matrix elements as $\rho$ on the central cross (up to normalisation), and all vanishing elements outside the central cross.
This is a resource destroying map \cite{liu2017resource} which removes any residual daemonicity from the state. Notice that this operation annihilates maximally daemonic states, ${\cal P}(\rho_{\max})=\mathbf{0}$, for which no corresponding angelic state can be defined. This can be seen as a form of {\em no-redemption theorem}. 

%%%%%%%%%%%%%%%%%%%%%%%%%%%%%%
\section{Resource quantifiers and state manipulation}

%%%%%%%%%%%%%%%%%%%%%%%%%%%%%
\subsection*{Wickedness}

It has widely been conjectured in the literature \cite{alighieri1909divine} that the Dante rank of a state should also be explicitly related to its degree of daemonic resource.
In this Section we show that this is indeed the case. 

Let $\Pi_{\mathcircled{\text{\tiny{H}}}_k}$ denote the projector onto the $k$th circle of Hell ($1 \leq k \leq 9$). Note that this operation acts as an exorcism if $k$ is smaller than the Dante rank of the state, but it can also create daemonic resource if $k$ is chosen to be larger (as a consequence of the free will theorem \cite{freewill}). For completeness, we can extend the definition to the case $k=0$ by setting  $\Pi_{\mathcircled{\text{\tiny{H}}}_0} \rho \Pi_{\mathcircled{\text{\tiny{H}}}_0} = {\cal P}(\rho)$, where the pinning operation ${\cal P}$ is defined in Eq.~\eqref{eq:pinning}. 

It's well known  \cite{bible00} that an entity cannot be found in a circle of Hell without possessing some wickedness. We therefore  define the {\em relative entropy of wickedness} of a state $\rho$ as the cumulative distance of the state from each circle of Hell,
\begin{equation}\label{wickedness}
    \mathcal{W}(\rho) = \sum\limits_{k=0}^9 \min\limits_{\sigma_k \in \mathcircled{\text{\tiny{H}}}_k} S(\rho || \sigma_k),
\end{equation}
where $S(\rho || \sigma) = \tr(\rho \log \rho - \rho \log \sigma)$ is the \emph{Kullback-Lucifer} divergence (with $\log$ taken in base $666$) and the minimization is performed over the set $\mathcircled{\text{\tiny{H}}}_k$ of all states with Dante rank $k$, with the identification $\mathcircled{\text{\tiny{H}}}_0 \equiv \mbox{\ding{63}}$.

%It is now straightforward to see that any state with non-zero projection onto at least one of the circles of Hell $\mathcircled{\text{\tiny{H}}}_k$ (beyond $k=0$), will necessarily admit a wicked decomposition. The proof of the converse, along with details of how to construct such a decomposition, are deferred to the Transcendental Material \cite{SupMatX}. 

Notice that the relative entropy of wickedness provides a less coarse-grained quantification of the resource content of daemonic states, as compared to the Dante rank. For a yet finer-grained account, one may instead consider the vector of elements $w_k(\rho)=\min\limits_{\sigma_k \in \mathcircled{\text{\tiny{H}}}_k} S(\rho || \sigma_k)$ characterizing the specific wickedness component of the state, with respect to each individual circle of Hell. This representation, which is similar in spirit to the characterization of multilevel  coherence in quantum information theory \cite{ringbauer2018certification}, will be referred to as  {\em Charon's order of wickedness} (COW) and proves useful as a tool to monitor the dynamics of the state towards its most representative circle of Hell. 
%to map the daemonic nature of the initial  quantum state, and to inform its dynamical navigation  towards a steady state belonging to its more representative circle of Hell (that is, the one $\mathcircled{\text{\tiny{H}}}_k$ indexed by the highest order $w_k$). 
Further applications of the COW to the study of physical and unphysical resources will be investigated in subsequent work.

%%%%%%%%%%%%%%%%%%%%%%%%%%%%%
\subsection*{Creation, sublimation, and demo-majorization}

We next show that the wickedness can also be used to define the concept of demo-majorization and hence to determine when specific transformations between daemonic forms, such as {\em shapeshifting} and {\em biomorphing}, are possible or not. To accomplish this, we first extend Eq.~(\ref{wickedness}) to the case of R\'eniy divergences by defining the R\'enyi-$\omega$ wickedness, 
\begin{equation}\label{wickedness_omega}
    \mathcal{W}_\omega(\rho) = \sum\limits_{k=0}^9 \min\limits_{\sigma_k \in \mathcircled{\text{\tiny{H}}}_k} S_\omega(\rho || \sigma_k).
\end{equation}
The use of R\'enyi-$\omega$ instead of R\'enyi-$\alpha$ divergences is quite revealing, as first shown in Ref.~\cite[John]{bible00}

First of all, it allows us to define two important operational measures, the {\em wickedness of creation} $\mathcal{W}_\infty(\rho)$ and the {\em wickedness of sublimation} $\mathcal{W}_0(\rho)$, which correspond to choosing $\omega=\infty$ and $\omega=0$ in Eq.~\eqref{wickedness_omega}, respectively. The wickedness of creation measures how many units of pure evil need to be sacrificed to create the state $\rho$, and the wickedness of sublimation quantifies how many units of pure evil can be eviscerated back from $\rho$ by means of exorcisms. While these definitions are valid in the single-shot regime, they can be naturally extended to the case of asymptotically many demons (a {\em legion}) \cite[Mark 5:9]{bible00}. 

Next, it is then straightforward to define the concept of {\em demo-majorization}, which we denote by the symbol $\xrightwitchonpitchfork{}$:
Given two density matrices $\rho$ and $\sigma$, 
\begin{equation}
\mathcal{W}_\omega(\rho) \geq \mathcal{W}_\omega(\sigma), \quad \forall \omega \qquad \Longleftrightarrow \qquad 
    \rho \xrightwitchonpitchfork{} \sigma.
\end{equation}
In this case we say that $\rho$ demo-majorizes $\sigma$.

The central result of this Section is that a daemonic state $\rho$ can be converted into another less daemonic state $\sigma$ by means of free exorcisms if and only if $\rho$ demo-majorizes $\sigma$.
The proof is provided in Section DCLXVI of the Transcendental Material~\cite{SupMatX}. This completely characterizes the power of manipulation of exorcisms on daemonic states in terms of the hierarchy of R\'enyi-$\omega$ wickedness measures.

An even stronger result concerns conversions and morphing between different wickedness traits within quantum states. This requires extending the concept of demo-majorization to a spiral of conditions on the COW structure of daemonic states. Due to the sensitive nature of this process, we are better left with leaving such an assumption unlifted in the present study.

%\begin{figure}
%    \centering
%    \includegraphics[width=7.5cm]{academicinferno.jpg}
%    \caption{Representation of the Hamiltonian operators $H_k$ in an academic projection of the nine circles of Hell. Source: \url{www.phdcomics.com}.}
%    \label{figdante}
%\end{figure}

%%%%%%%%%%%%%%%%%%%%%%%%%%%%%%
\section{Connection with the resource theory of thermodynamics}

%%%%%%%%%%%%%%%%%%%%%%%%%%%%%

We finally hereby provide a consistent construction of a resource theory of thermodynamics beyond the realm of the physical world, exactly \textit{in the same spirit} of the previous Sections. We remark that one could construct further generalizations of our approach, e.g., to the resource theory of (black) magic, (ethical) coherence, (spiritual) asymmetry, and so on. However, these will be explored in future works, and the remainder of this paper shall be mainly devoted to unveil the thermodynamical implications of our results.

While originally conceived to describe the change in energy and entropy of macroscopic engines, thermodynamics has been successfully applied to meso- and nano-scale systems. This research field, named quantum thermodynamics \cite{goold2016role,alicki2018introduction}, has led to many breakthrough results and refinements of the usual three laws of thermodynamics, valid at the level of fluctuating quantities. Not enough attention has yet been given however to the important class of ethereal states considered in this work and their power for work and heat exchange. We will show that this new angle will also allow us to shed light on the famous Maxwell's demon paradox.

First of all, as done in Section \ref{sec:construction}, we proceed to define the set of free states in our resource theory of (supernatural) thermodynamics. The latter are defined as the states $\tau_{k} \in \mathcircled{\text{\scriptsize{H}}}_k $ of the form 
\begin{equation}\label{kDemon}
    \tau_{k} = \frac{\exp\left[-\beta_{k} H_k\right]}{Z_k},
\end{equation}
where $\beta_k$ is the inverse temperature, $Z_k$ denotes the partition function, and $H_k$ stands for the Hamiltonian operator in the $k$th circle of Hell. % For a visual representation of such operators, see Fig.~\ref{figdante}.
For each $\mathcircled{\text{\scriptsize{H}}}_k$, $k = 1, \ldots, 9$, $\tau_{k}$
represents a daemonic state perfectly thermalized within the infernal environment of its circle of Hell. We call any such state $\tau_{k}$ a \textit{$k$-Maxwell's demon}.

As first pointed out in \cite{alighieri1909inferno}, subsequent circles of Hell are characterized by a non-decreasing temperature, i.e. $\beta_{k} \geq \beta_{k+1}$, for $k = 1, \ldots , 8$; the 9-Maxwell's demon, however, is found to be at zero temperature, i.e.~$\beta_{9} = +\infty$. It is also worth pointing out that this fixes the state $\tau_9$ to be a pure (evil) state, which can be shown to actually correspond to the maximally daemonic state $\rho_{max}$, as introduced in Eq.~\eqref{maxdoom}, in dimension $9j+1$ ($\forall \ j \geq 2$).
This dramatic inversion of temperature trend will have important thermodynamic consequences, as shown below. 

The set of free operations in this setting, called \textit{guilt-preserving} (GP) operations, which contain exorcisms as a subset, also  preserve the $k$-Maxwell's demons, i.e. 
\begin{equation}\label{SPmaps}
    \mathcal{E}_{GP}\left(\rho\right) = \sum_k \tr_{\mathcircled{\text{\scriptsize{H}}}_k}\bigg\{ U ( \rho \otimes \tau_{k}) U^\dagger\bigg\}.
\end{equation}
Equipped with these notions and Eq. \eqref{SPmaps}, one can wonder, given two states $\rho$ and $\sigma$, whether they are connected through a GP operation. Remarkably, the answer is provided by a variant of the previously introduced notion of demo-majorization, which we call \textit{guilt-majorization} and denote with the symbol $\mathbat$:
given two states $\rho$ and $\sigma$, 
\begin{equation}
\mathcal{G}_\omega(\rho) \geq \mathcal{G}_\omega(\sigma), \quad \forall \omega \qquad \Longleftrightarrow \qquad 
    \rho \,\,\mathbat\,\, \sigma,
\end{equation}
where we introduced the \textit{guilt} quantifier
\begin{equation}
    \mathcal{G}_\omega(\rho) = \sum\limits_{k=1}^9 \min\limits_{\tau_{k} \in \mathcircled{\text{\scriptsize{H}}}_k} S_\omega(\rho || \tau_{k}).
\end{equation}
In the Transcendental Material \cite{SupMatX} we demonstrate that  guilt-majorization represents a necessary and sufficient condition for this single-shot transformation to be allowed, i.e. 
\begin{equation}
\rho \longrightarrow \sigma = \mathcal{E}_{GP}\left(\rho\right)  \quad \Longleftrightarrow \quad  \rho \,\, \mathbat \,\, \sigma.
\end{equation}

Borrowing from the existing literature on resource theory of thermodynamics \cite{goold2016role}, one can then enlarge the set of reachable states by making use of an ancilla $\lambda_B$, controlled by an additional party Beelzebub --- hereby referred to as the {\em devil's advocate} \cite{devilsadvocate} --- through the following catalytic transformation
\begin{equation}
\mathcal{E}_{GP}(\rho)\otimes \lambda_B = \tr_{\mathcircled{\text{\scriptsize{H}}}_k}\bigg\{ U ( \rho \otimes \lambda_B \otimes \tau_{k}) U^\dagger\bigg\}.
\end{equation}
Remarkably, we notice that if Beelzebub initializes the ancilla system $B$ in a state $\lambda_B$ chosen from the class of Black Schr\"{o}dinger cat states, which include in particular the DOOM states of Eq.~(\ref{maxdoom}), then one can obtain {\em cataclysmic transformations} of any state $\rho$ into any other daemonic state $\sigma$ by means of guilt-preserving operations.  This phenomenon will pave the way for potentially revolutionary applications to biological, theological, and environmental sciences  \cite{LunarCataclysm}.
We have a very elegant proof of this statement, but unfortunately it does not fit into our own circle of academic Hell and will be reported elsewhere (confession pending).

We conclude this Section by applying the above formalism to the study of the famous Maxwell's demon paradox, that is the motivation of this paper. 
As a consequence of Landauer's erasure principle, it is easy to see that a $k$-Maxwell's demon, as defined in Eq.~\eqref{kDemon}, must invest at least $ \Delta W_{diss} = \beta_{k} \log_{666} 2$ of work in order to increment the entropy of its circle of Hell by one dbit (daemonic bit). However, using the same arguments, it is straightforward to prove that, being at zero temperature, a $k$-Maxwell's demon with $k=9$ is able to violate the second law of (supernatural) thermodynamics at no expenditure of dissipated heat, thus representing the ideal realization Maxwell had envisioned in his seminal work \cite{Maxwell1888}. This can unlock the gates to a hidden Garden of Eden, thriving with unprecedented applications, both theoretical and experimental, beyond the realm of humanely imaginable knowledge \cite{miltonparadise}. 

Pursuing this path to discovery will no doubt require unfamiliar and uncountable resources. Our work is a first step towards identifying, benchmarking, exploiting, and making sense of these resources in everyday life and afterlife.

%%%%%%%%%%%%%%%%%%%%%%%%%%%%%%
\section{Conclusions}

%In this paper we have put forth a resource theory of Maxwell's demon. The theory aimed at addressing several open questions concerning the connection between information and thermodynamics. 

{\bf This article was published on April 1st.  Some of the results presented in this paper may not be fully supported by the quantum physics community. It is unclear whether the authors were possessed  during the preparation of this manuscript. But they are OK now. \\
}

{\em ``The mind is its own place, and in itself can make a Heaven of Hell, a Hell of Heaven''} \cite{miltonparadise}.
\\[0.3cm]

%%%%%%%%%%%%%%%%%%%%%%%%%%%%%
\subsection*{Acknowledgements}
The authors acknowledge fruitful discussions with D. Brown, A. Hathaway, Buffy, M. L. Ciccone, J. Milton, A. Pacino, P. Quemedo, A. Virmond, C. Sidoli, R. Maia, M. T.-C. Her, D. Bracchi, M. Teixeira, F. do Capeta, Z. do Caix\~ao, H. Asbro, P. Marcelo, P. P. G. Him, and Rita from Praia do Madeiro.  The authors further thank M. Paternostro and the R. Stones for inspiration. The authors acknowledge the Anhanguera Institute for Niobium-based Quantum Technologies and the International Institute of Christmas, where parts of this work were developed, for both the hospitality and the psychological support. 

%\bibliography{/Users/gtlandi/Documents/library}
\bibliographystyle{ieeetr}
\bibliography{library,Index}

\begin{thebibliography}{10}

\bibitem{Maxwell1888}
J.~C. Maxwell, {\em {The theory of heat}}.
\newblock Longmans, Green, and Company, New York, 1888.

\bibitem{Toyabe2010}
S.~Toyabe, T.~Sagawa, M.~Ueda, E.~Muneyuki, and M.~Sano, ``{Experimental
  demonstration of information-to-energy conversion and validation of the
  generalized Jarzynski equality},'' {\em Nature Physics}, vol.~6, p.~988,
  2010.

\bibitem{Koski2014}
J.~V. Koski, V.~F. Maisi, J.~P. Pekola, and D.~V. Averin, ``{Experimental
  realization of a Szilard engine with a single electron},'' {\em Proceedings
  of the National Academy of Science}, vol.~111, p.~13786, 2014.

\bibitem{Camati2016}
P.~A. Camati, J.~P.~S. Peterson, T.~B. Batalh{\~{a}}o, K.~Micadei, A.~M. Souza,
  R.~S. Sarthour, I.~S. Oliveira, and R.~M. Serra, ``{Experimental
  Rectification of Entropy Production by Maxwell's Demon in a Quantum
  System},'' {\em Physical Review Letters}, vol.~117, no.~24, p.~240502, 2016.

\bibitem{Masuyama2017b}
Y.~Masuyama, K.~Funo, Y.~Murashita, A.~Noguchi, S.~Kono, Y.~Tabuchi,
  R.~Yamazaki, M.~Ueda, and Y.~Nakamura, ``{Information-to-work conversion by
  Maxwell's demon in a superconducting circuit-QED system},'' {\em Nature
  Communications}, vol.~9, no.~1291, pp.~1--16, 2018.

\bibitem{Cottet2017}
N.~Cottet, S.~Jezouin, L.~Bretheau, P.~Campagne-Ibarcq, Q.~Ficheux, J.~Anders,
  A.~Auff{\`{e}}ves, R.~Azouit, P.~Rouchon, and B.~Huard, ``{Observing a
  quantum Maxwell demon at work},'' {\em Proceedings of the National Academy of
  Science}, vol.~114, no.~29, pp.~7561--7564, 2017.

\bibitem{Xiong2018}
T.~P. Xiong, L.~L. Yan, F.~Zhou, K.~Rehan, D.~F. Liang, L.~Chen, W.~L. Yang,
  Z.~H. Ma, M.~Feng, and V.~Vedral, ``{Experimental Verification of a
  Jarzynski-Related Information-Theoretic Equality by a Single Trapped Ion},''
  {\em Physical Review Letters}, vol.~120, no.~1, p.~010601, 2018.

\bibitem{goold2016role}
J.~Goold, M.~Huber, A.~Riera, L.~del Rio, and P.~Skrzypczyk, ``The role of
  quantum information in thermodynamics --- a topical review,'' {\em Journal of
  Physics A: Mathematical and Theoretical}, vol.~49, no.~14, p.~143001, 2016.

\bibitem{laplacedemon}
P.~S. Laplace, ``A philosophical essay on probabilities, 1819,'' {\em English
  translation, Dover}, vol.~6, 1951.

\bibitem{Micadei2017b}
K.~Micadei, J.~P.~S. Peterson, A.~M. Souza, R.~S. Sarthour, I.~S. Oliveira,
  G.~T. Landi, T.~B. Batalh{\~{a}}o, R.~M. Serra, and E.~Lutz, ``{Reversing the
  thermodynamic arrow of time using quantum correlations},'' {\em arXiv
  1711.03323}, 2017.

\bibitem{ouija}
A.~R. Feinstein, ``Xxxix. the haze of bayes, the aerial palaces of decision
  analysis, and the computerized ouija board,'' {\em Clinical Pharmacology \&
  Therapeutics}, vol.~21, no.~4, pp.~482--496.

\bibitem{LunarCataclysm}
F.~Tera, D.~Papanastassiou, and G.~Wasserburg, ``Isotopic evidence for a
  terminal lunar cataclysm,'' {\em Earth and Planetary Science Letters},
  vol.~22, no.~1, pp.~1 -- 21, 1974.

\bibitem{Guarnieri2017}
G.~Guarnieri, S.~Campbell, J.~Goold, S.~Pigeon, B.~Vacchini, and
  M.~Paternostro, ``{Full counting statistics approach to the quantum
  non-equilibrium Landauer bound},'' {\em New Journal of Physics}, vol.~19,
  no.~10, pp.~1--9, 2017.

\bibitem{SupMatX}
See Transcendental Material (acessible via the Deepweb).

\bibitem{alighieri1909inferno}
D.~Alighieri, {\em Dante's Inferno}.
\newblock Classic Dante, 1320.

\bibitem{wikipedia}
AnonymousDeepWebUser, ``Chebyshev distance,'' {\em Wikipedia}, 1355.

\bibitem{exorcist}
W.~P. Blatty, {\em {The Exorcist}}.
\newblock Harper and Row, USA, 1971.

\bibitem{DevilWearsPrada}
L.~Weisberger, {\em {The Devil Wears Prada}}.
\newblock Broadway Books, USA, 2004.

\bibitem{liu2017resource}
Z.-W. Liu, X.~Hu, and S.~Lloyd, ``Resource destroying maps,'' {\em Physical
  Review Letters}, vol.~118, no.~6, p.~060502, 2017.

\bibitem{alighieri1909divine}
D.~Alighieri, {\em The Divine Comedy}.
\newblock Classic Dante, 1320.

\bibitem{freewill}
J.~Conway and S.~Kochen, ``The free will theorem,'' {\em Foundations of
  Physics}, vol.~36, no.~10, pp.~1441--1473, 2006.

\bibitem{bible00}
God, {\em The Bible}.
\newblock King James, 1611.

\bibitem{ringbauer2018certification}
M.~Ringbauer, T.~R. Bromley, M.~Cianciaruso, L.~Lami, W.~S. Lau, G.~Adesso,
  A.~G. White, A.~Fedrizzi, and M.~Piani, ``Certification and quantification of
  multilevel quantum coherence,'' {\em Physical Review X}, vol.~8, no.~4,
  p.~041007, 2018.

\bibitem{alicki2018introduction}
R.~Alicki and R.~Kosloff, ``Introduction to quantum thermodynamics: History and
  prospects,'' {\em arXiv preprint arXiv:1801.08314}, 2018.

\bibitem{devilsadvocate}
{\em The Devil's Advocate} directed by Taylor Hackford, 1997.

\bibitem{miltonparadise}
J.~Milton, {\em Paradise Lost}.
\newblock Hackett Publishing, 1667.

\end{thebibliography}
%\end{document}

\end{document}